\documentclass[twocolumn]{aastex63}

\usepackage{bm}
\usepackage{amsmath}

\received{April 29, 2021}
\revised{\today}

\submitjournal{ApJ}

\shorttitle{Bayesian Inference of GC Properties Using Distribution Functions}
\shortauthors{Eadie, Webb, and Rosenthal}

\begin{document}

\title{Bayesian Inference of Globular Cluster Properties Using Distribution Functions}

\correspondingauthor{Gwendolyn M. Eadie}
\email{gwen.eadie@utoronto.ca}

\author[0000-0003-3734-8177]{Gwendolyn M. Eadie}
\affiliation{David A. Dunlap Department of Astronomy \& Astrophysics,University of Toronto, Toronto, ON}
\affiliation{Department of Statistical Sciences, University of Toronto, Toronto, ON}

\author[0000-0003-3613-0854]{Jeremy J. Webb}
\affiliation{David A. Dunlap Department of Astronomy \& Astrophysics,University of Toronto, Toronto, ON}

\author[0000-0002-5118-6808]{Jeffrey S. Rosenthal}
\affiliation{Department of Statistical Sciences, University of Toronto, Toronto, ON}

\begin{abstract}

We present a Bayesian inference approach to estimating the cumulative mass profile and mean squared velocity profile of a globular cluster given the spatial and kinematic information of its stars. Mock globular clusters with a range of sizes and concentrations are generated from lowered isothermal dynamical models, from which we test the reliability of the Bayesian method to estimate model parameters through repeated statistical simulation. We find that given unbiased star samples, we are able to reconstruct the cluster parameters used to generate the mock cluster and the cluster's cumulative mass and mean velocity squared profiles with good accuracy. We further explore how strongly biased sampling, which could be the result of observing constraints, may affect this approach. Our tests indicate that if we instead have biased samples, then our estimates can be off in certain ways that are dependent on cluster morphology. Overall, our findings motivate obtaining samples of stars that are as unbiased as possible. This may be achieved by combining information from multiple telescopes (e.g., \textit{Hubble} and \textit{Gaia}), but will require careful modeling of the measurement uncertainties through a hierarchical model, which we plan to pursue in future work.

\end{abstract}


\section{Introduction}\label{sec:intro}

Globular clusters are nearly-spherical, massive collections of stars that are found in every type of galaxy. Upon formation, their early evolution is governed by stellar evolution in the sense that massive stars quickly lose mass, which causes the cluster's potential to weaken. However, over the majority of their lifetimes, two-body relaxation and the external tidal field of their host galaxy are the dominant mechanisms that govern a cluster's evolution \citep[e.g.][]{Heggie2003}. These two mechanisms lead to clusters becoming spherically symmetric, isotropic, and mass segregated over time as they evolve towards a state of partial energy equipartition while playing host to stellar collisions and mergers \citep{Meylan1997, Spitzer1987, Heggie2003}. Dynamically old clusters are even capable of having their core energetically decouple from the rest of the cluster, a process known as core collapse \citep{Henon1961,LyndenBell1968}.

Given the bevy of dynamical processes that occur within globular clusters, the ability to accurately measure the current distribution of stars within a given cluster leads to a deeper understanding of how these processes work and shape cluster evolution. Reverse engineering the evolution of a system of clusters can then lead to constraining the conditions under which they form and therefore the formation and evolution of their host galaxy. A large number of distribution functions (DFs) have been proposed to represent the observed distribution of stellar positions and velocities in globular clusters \citep[e.g.,][]{Woolley1954, Michie1963, King1966, Wilson1975, Gunn1979, Bertin08, Gieles2015, Claydon2019}. The general picture that emerges out of the models that best represent observations of Galactic globular clusters is that clusters are isotropic in their centre with density and velocity dispersion profiles that decrease to zero out to a truncation radius. The treatment of how the DF drops to zero out to the truncation radius varies from model to model, with additional treatments being necessary to address the presence of radial anisotropy \citep{Michie1963} and globular cluster rotation \citep{Varri2012}. 

Complicating the situation slightly is that stars within globular clusters have a large range of masses, while most DFs assume all stars have the same mass. Hence mass segregation, which is a natural outcome of clusters evolving towards a state of partial energy equipartition, is not considered in the models. Failing to account for the presence of mass segregation has been shown to incur strong biases when fitting models to the surface brightness profile or number density profile of a cluster \citep{Shanahan2015, Sollima2015}. One solution is to treat a globular cluster system as the combination of several single mass models \citep{DaCosta1976}. 

Historically, the application of the aforementioned models to observed globular clusters has been in the fitting of their observed number density or surface brightness profiles. From a given distribution function, it is possible to derive how the number of stars per unit area on the sky or volume decreases with clustercentric distance. Assuming a mass spectrum and mass-to-light ratio, a surface brightness profile can also be derived. Several different distribution function-based models have been successfully fit to Galactic \citep[][e.g]{McLaughlin2005,Miocchi2013, deBoer2019} and extragalactic \citep[][e.g]{Woodley2010, Usher2013, Webb2013, Puzia2014} globular clusters. 

Alternatives to fitting clusters with distribution function based models include comparing observations to large suites of $N$-body star cluster simulations \citep{Heggie2014, Baumgardt2018} and Jeans Modelling \citep{Cappellari2008, Watkins2013}. Direct $N$-body simulations can also be used to test and rule out different distribution function based models, as completeness, contamination and measurement errors will not contribute to the uncertainty in the fit. For example, \citet{Zocchi2016} successfully demonstrated that direct $N$-body simulations of star clusters could be well fit by the lowered isothermal models of \citet{Gieles2015}.

In addition to the issues associated with assuming what model best represents globular clusters in general, the process of finding the exact model parameters (or $N$-body simulation) that best represent a specific globular cluster is also challenging. Historically, globular clusters were fit with models by comparing observed and theoretical surface brightness profiles or density profiles \citep[e.g,][]{McLaughlin2005}. A typical approach to fitting observational data with models would be to radially bin the observed stars and then minimize the $\chi^2$ between the observed surface brightness or density profile and the model profile. Such an approach will result in systematic error due to binning the data, with the completeness of the dataset, contamination from non-cluster stars, and measurement errors introducing additional uncertainty into the fit as well. Binning data is also undesirable as information is lost about each individual star. Furthermore, as previously mentioned, multi-mass models require either a  mass-to-light ratio be added as a free parameter when fitting surface brightness profiles or a mass-to-light ratio be assumed for the observational data \citep{2019MNRAS.483.1400H}.

Gaia Data Release 2 \citep{Gaia16a,gaia18} and the Hubble Space Telescope Proper Motion (HSTPROMO) Survey \citep{Bellini2014} have helped usher in a new era of globular cluster studies, with spatial and kinematic information now available for a large number of cluster stars. Knowing the kinematic properties of individual stars can mitigate uncertainties related to contamination, as kinematics make it easier to determine what stars in the observed field of view are truly members of the cluster or are simply foreground or background stars. Combining membership constraints with spatial and photometric information of core stars in high-resolution images of cluster centres also allows for the radial coverage across a cluster to be improved \citep{deBoer2019}. 

Kinematic information can also be taken into consideration when fitting clusters with models, as the cluster's density profile and velocity dispersion can be simultaneously fit by minimizing the combined $\chi^2$ \citep{Baumgardt2018}. Extending the method even further, \cite{Zocchi2017} has fit lowered isothermal models to the Galactic globular cluster Omega Centauri by simultaneously fitting its surface brightness profile, line of sight velocity dispersion profile, radial proper motion dispersion profile, and tangential proper motion dispersion profile. Unfortunately, even with kinematic information, issues related to binning data, completeness, and measurement uncertainties remain when fitting data with models. Furthermore, when trying to simultaneously fit surface brightness profiles and kinematic profiles, one must assume how to weight the importance of each fit. For example, when fitting through the minimization of $\chi^2$ between model and observed data, it must be decided whether the total $\chi^2$ is simply the sum of the individual $\chi^2$ values calculated for the density and kinematic profile fits or if they should be weighted differently. The advantages and disadvantages of fitting each of the models discussed above to observed cluster datasets are summarized by \citet{2019MNRAS.483.1400H}.

The purpose of this study is to investigate and potentially improve the method in which distribution function-based models can be fit to observed star cluster datasets by avoiding systematic errors and loss of information associated with radially binning the data, contamination, and completeness. We instead estimate the model parameters, cumulative mass profile, and mean-square velocity profile of a globular cluster (GC) using the positions and velocities of individual stars and assuming a physical model for the GC through a DF and Bayesian method. 

A Bayesian framework has at least four main advantages for this type of analysis. First, we wish to incorporate useful prior information about GCs to help constrain parameter estimates. Second, since kinematic data for GCs is often incomplete, using a Bayesian framework allows one to include both incomplete and complete data simultaneously. Third, astronomical data are also subject to measurement uncertainties that are well understood by astronomers, and that we can incorporate via a hierarchical Bayesian framework. Fourth, our ultimate goal is to infer the cumulative mass profile without having to make assumptions about the mass-to-light ratio of the GC, and this should be achievable given samples from the posterior distribution of model parameters. 

For the current study, we work with simulated data generated using \texttt{limepy} \citep{Gieles2015} of lowered isothermal models for GCs and test the ability of a Bayesian framework to recover a cluster's true total mass, cumulative mass profile, mean-square velicity profile, and other parameters of interest. A related study was completed by \citet{2019MNRAS.483.1400H}, where they used a single snapshot from a direct $N$-body simulation of the Galactic GC M4 \citep{Heggie2014} to compare the ability of multiple \textit{methods} to recover the simulated cluster's mass and mass profile. In the current paper, rather than comparing and contrasting the pros and cons of different methodological approaches on a single snapshot, we study the pros and cons of a single method to recover the mass profile of different types of of globular clusters (e.g., ``average'', ``compact'',  ``extended'' GCs). This approach is especially important, as \cite{2019MNRAS.483.1400H} suggested that single-mass DF methods could lead to biases in the mass and mass profile. We would like to concretely quantify any possible biases, and identify whether they are dependent on certain types of GCs (e.g., average, compact, and extended).

The paper is structured as follows. In Section \ref{sec:data}, we introduce the suite of simulated data used to test our approach, with the fitting routine and methods described in Section \ref{sec:methods}. In Section \ref{sec:results}, we examine the estimated coverage probabilities of the Bayesian credible intervals for the model parameters, and discuss situations in which inference from the posterior distribution is (and is not) able to reproduce the true cumulative mass profile of the simulated GCs. Future applications of this work, including the use of observational data, are also discussed. Finally, we summarize our findings in Section \ref{sec:conclusion}.

\section{Simulated Data}\label{sec:data}

We develop and test our method for GC parameter inference with simulated kinematic data $\bm{d} = (\mathbf{r}, \mathbf{v})$ of stars in a GC-centric reference frame, where $r_i = \sqrt{x_i^2 + y_i^2 + z_i^2}$ and $v_i= \sqrt{v_{x,i}^2 + v_{y,i}^2 + v_{z,i}^2}$ are the distance and speed of the $i^{th}$ star. The data are generated using the python code \texttt{limepy} \citep{Gieles2015}, which uses a five-parameter model for the phase-space distribution function $f(\bm{r},\bm{v})$ of stars in the cluster (see Section~\ref{sec:methods}). The \texttt{limepy} parameters are
\begin{eqnarray}
    \bm{\theta_{\text{limepy}}} & = & (g, \Phi_0, M, r_h, r_a)
    \label{eq:pars}
\end{eqnarray}
where $g$ (dimensionless) is a truncation parameter, $\Phi_0$ (dimensionless) determines the central potential, $M_{total}$ (in $M_{\odot}$) is the total mass, $r_h$ (in parsecs (pc) is the half-light radius, and $r_a$(pc) is the anisotropic radius of the GC. In this work we focus only on isotropic GCs, i.e., $r_a\rightarrow\infty$ (the default in \texttt{limepy}). Overall, $g$ and $\Phi_0$ impact the shape of the GC profile, while $M_{total}$ and $r_h$ are scale parameters. In the case of isotropic GCs, a value of $g=0$ in the \texttt{limepy} model is equivalent to the \cite{Woolley1954} model, and a value of $g=1$ is quivalent to the King models (\citealt{Michie1963,King1966}, see also \citealt{Gieles2015}). 

The combination of parameter values $g$, $\Phi_0$ and $r_h$ together determine both the  ``compactness'' and concentration of the GC. The value of $g$ is not only a truncation parameter but also plays a role in determining the spatial distribution of stars. The parameter $\Phi_0$ --- which determines the central gravitational potential --- helps set the concentration of stars. At the same time, GCs with the the same $M_{total}$, $g$, and $\Phi_0$, but with different half-light radii $r_h$, will also have different relative ``compactness''. Thus, it is the combination of these parameters that determine the GC morphology.

In this work, we explore different GC morphologies based on the parameter values listed in Table~\ref{tab:partable}. Every simulated GC has the same total mass ($M_{total}=10^5M_{\odot}$) and truncation parameter $g=1.5$, but have different levels of ``compactness'' (different $r_h$) or different concentrations (different $\Phi_0$). When varying $r_h$, we refer to GCs with $r_h=1.0$ as ``compact'', $r_h=3.0$ as ``average'', and $r_h=9$ as ``extended''. When varying $\Phi_0$, we refer to GCS with $\Phi_0=2$ as ``Low $\Phi_0$'', $\Phi_0=5$ as ``average'' (the same as the GC generated when $r_h=5.0$), and  $\Phi_0=9.0$ as ``High $\Phi_0$``. Thus, these combinations provide five scenarios: ``compact'', ``average'', ``extended'', ``High $\Phi_0$'', and ``Low $\Phi_0$''. We create 50 GCs of each type in order to repeat our analysis many times.

\begin{figure*}
    \centering
    \includegraphics[scale=0.6]{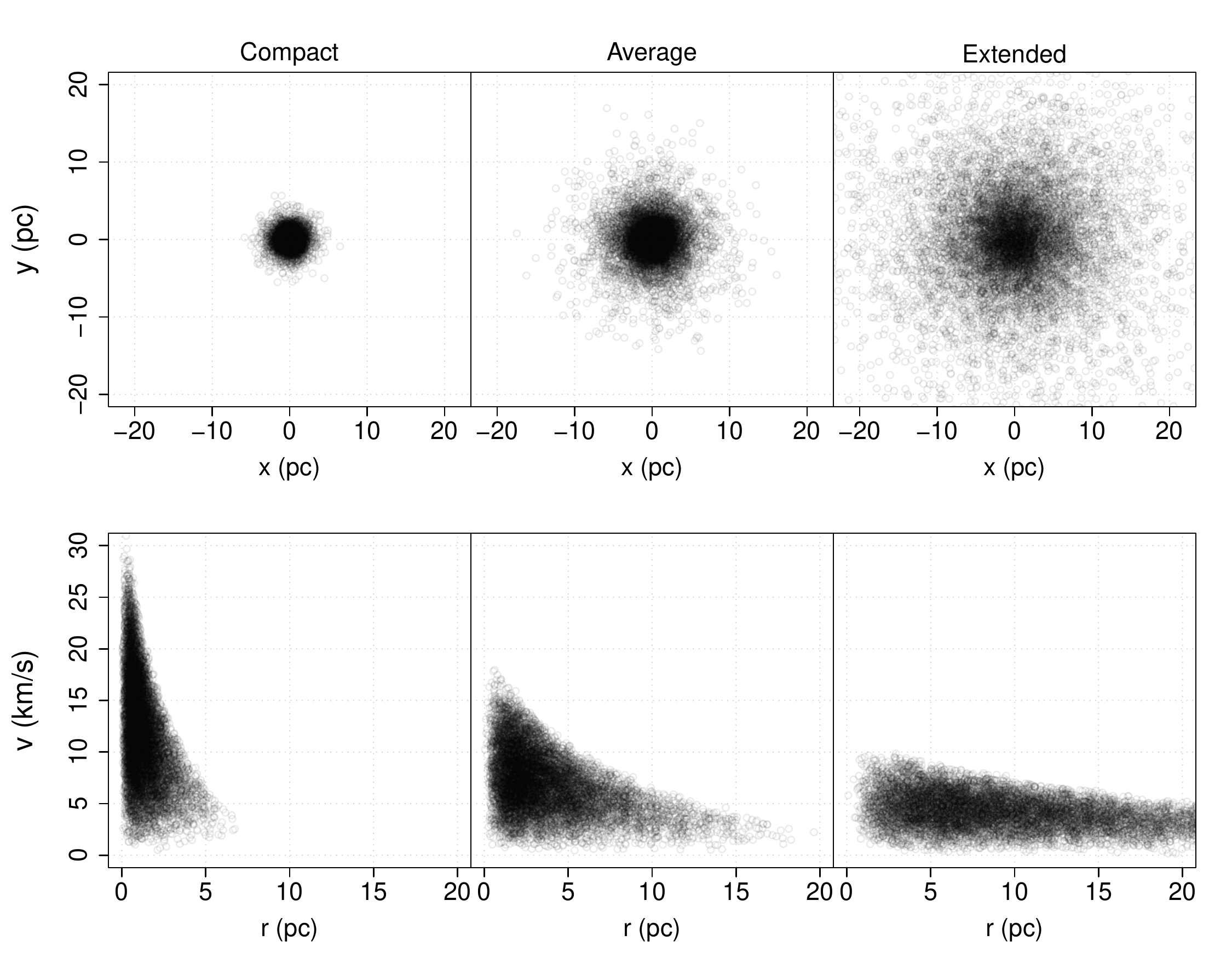}
    \includegraphics[scale=0.6]{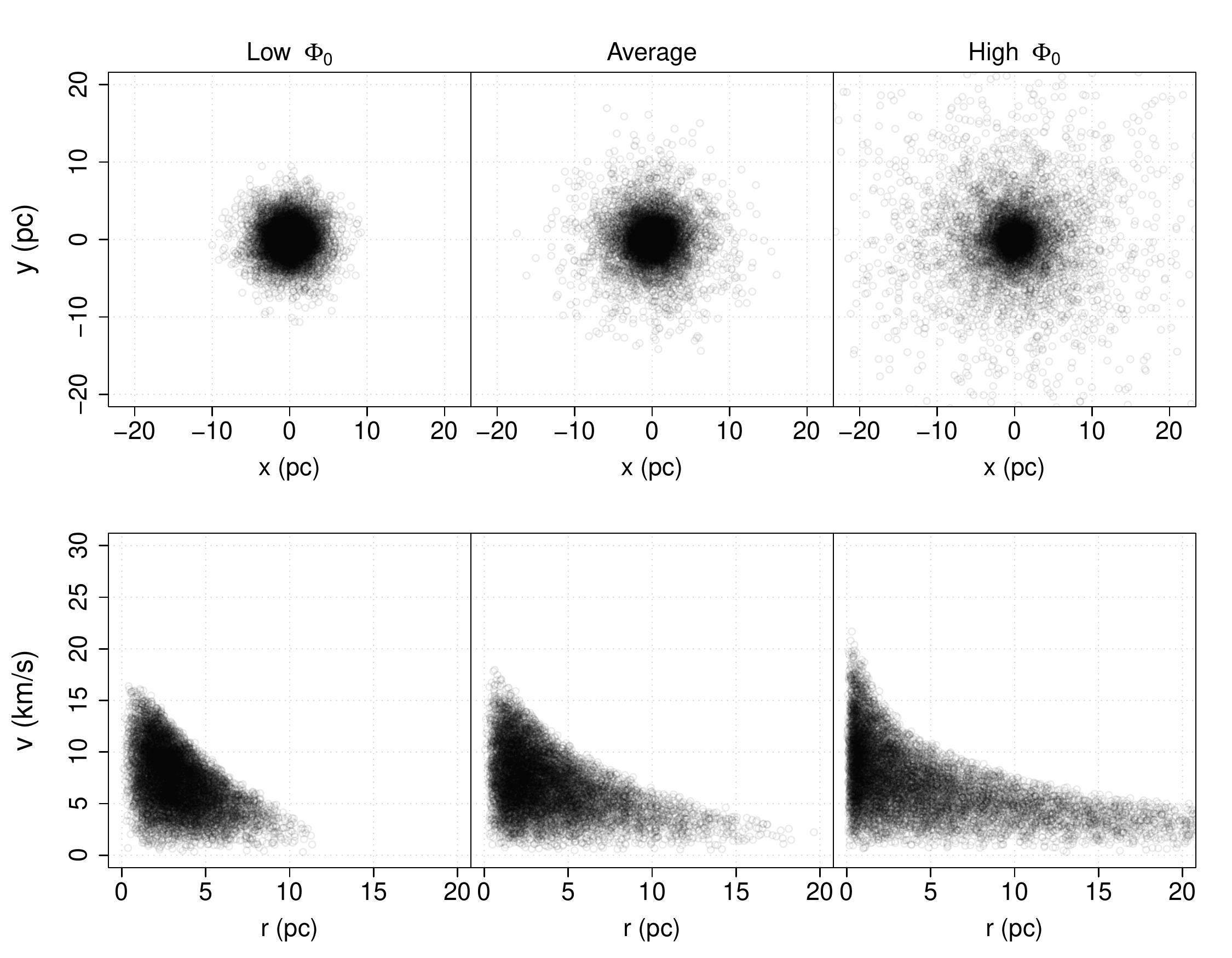}
    \caption{\textit{First row:} The $x$ and $y$ coordinates of ten thousand randomly selected stars in three different simulated GCs: a compact (left), average (middle), and extended (right) GC with $10^5$ stars and \texttt{limepy} parameters $g=1.5$, $\Phi_0 = 5.0$, $M=10^5$, and $r_h = 1.0, 3.0$ and $9.0$ respectively. \textit{Second row:} The total velocity profile as a function of total distance $r$, using the same stars as in the top row. \textit{Third and fourth rows:} The same as the top two rows, except for a GC with parameters $g=1.5$, $M=10^5$, and $r_h = 3.0$, and changing the $\Phi_0$ parameter: $\Phi_0=2.0$ (left), average (middle, same as top two rows), and $\Phi_0=8.0$.}
    \label{fig:3GCs}
\end{figure*}

Figure~\ref{fig:3GCs} shows examples of the $x$ and $y$ positions of GC stars (first and third rows) and their absolute velocity profiles (second and fourth rows) in clustercentric coordinates. These kinematic data were created using $g=1.5$, $\Phi_0 = 5.0$, $M=10^5$, and $r_h = 1.0, 3.0$ and $9.0$ respectively for the top two rows, and using $g=1.5$, $M=10^5$, $r_h=3$, and $\Phi_0 = 2.0, 5.0$,  and $\Phi_0 = 8.0$ respectively for the bottom two rows. The ``average'' GC is shown in all rows to show the transition from low $r_h$ to high $r_h$ and from low $\Phi_0$ to high $\Phi_0$.

\begin{table}
    \centering
    \begin{tabular}{l|l|c}
        \textbf{$\bm{\theta}$} & \textbf{Description} & \textbf{Possible values} \\
        \hline
         $g$ & truncation parameter & 1.5 \\
         $\Phi_0$ & central gravitational potential & 2.0, 5.0, 8.0 \\
         $M_{total}$ & total mass ($M_{\odot}$)  & $10^5$ \\
         $r_h$ & half-light radius (pc) &   1.0, 3.0, 9.0 \\
    \end{tabular}
    \caption{\texttt{limepy} model parameter values used to simulate stars in GCs. }
    \label{tab:partable}
\end{table}

Each simulated GC contains $n=10^5$ stars. In real data sets, we do not have kinematic information for all $n$ stars due to limited observations and observational selection effects. Thus, we study the effects of our mass profile estimates when selecting stars (a) randomly, (b) only in the outer regions (thereby mimicking \textit{Gaia} data), and (c) only in the inner regions (thereby mimicking \textit{HST} data). In each case, we use a subsample of 500 stars from each GC. Moreover, these three different tests, combined with the five different morphological GCs (compact, average, extended, low $\Phi_0$, and high $\Phi_0$), leads to fifteen different scenarios (Table~\ref{tab:cases}).

For this initial study and for the development and testing of our code, we use complete data in both position and velocity and assume there is no measurement uncertainty. We also work in the reference frame of the GC, where positions and velocities of individual stars are given with respect to the GC center. Of course, real data are collected in a Heliocentric reference frame, may be incomplete (e.g., only projected distances and line-of-sight velocities are known), and are subject to measurement uncertainty. However, it is worthwhile to investigate the ability of this method in an idealized case where we have complete data. Ultimately, our goal is to work in projected space on the plane of the sky (i.e., the reference frame in which actual data are measured), account for incomplete data (e.g., when only one component of the position is known), and incorporate measurement uncertainty through a hierarchical model.

\section{Methods}\label{sec:methods}

Using the simulated spatial and kinematic data of stars from each GC mentioned in Section~\ref{sec:data}, we take a Bayesian approach to infer the model parameters of each GC. From Bayes' theorem \citep{bayes1763}, the posterior probability of a vector of model parameters $\bm{\theta}=(g, \Phi_0, M_{total}, r_h)$, given data $\bm{d}$, is
\begin{equation}
    p(\bm{\theta}|\bm{d}) = \frac{p(\bm{d}|\bm{\theta})p(\bm{\theta})}{p(\bm{d})},
    \label{eq:bayes}
\end{equation}
where $p(\bm{d}|\bm{\theta})$ is the probability of the data conditional on the model parameters, $p(\bm{\theta})$ is the prior distribution on the model parameters, and $p(\bm{d})$ is the ``evidence'' or \textit{prior predictive density}. The latter is a constant, leaving us with a target distribution proportional to the posterior distribution $p(\bm{\theta}|\bm{d})$, which we will estimate through sampling in order to perform parameter inference (Section~\ref{sec:mcmc}). Our simulated data $\bm{d}$ described in Section~\ref{sec:data} are the six Cartesian phase-space components  {$(\bm{x,y,z,v_x,v_y,v_z)}$} of each star, which we treat as perfectly measured. An individual star's phase-space components $d_i=(x_i,y_i,z_i,v_{x,i},v_{y,i},v_{z,i})$ provide its clustercentric distance $r_i$ and speed $v_i$, which are needed for the calculation of the DF $f(\bm{\theta}; d_i)$. 

In practice, $p(\bm{d}|\bm{\theta})$ is often taken to be the likelihood --- a function of model parameters for fixed data  $\mathcal{L}(\bm{\theta};\bm{d})$ --- which we define using the DF in Section~\ref{sec:likelihood}. The prior distributions for the model parameters $\bm{\theta}=(g, \Phi_0, M_{total}, r_h)$ in the \texttt{limepy} model are described in Section~\ref{sec:priors}.

\subsection{Likelihood}\label{sec:likelihood}

In this study, we define the likelihood using a physical distribution function (DF), $f(\bm{\theta}; d_i)$ of the \texttt{limepy} lowered-isothermal model. Given a fixed set of data $\bm{d}$ of $N$ stars, the likelihood is a function of the model parameters $\bm{\theta}$ and the total mass $M_{total}$ of the GC:
\begin{eqnarray}
\mathcal{L}(\bm{\theta};\bm{d}) &=& \prod\limits_{i=1}^{N} \frac{f(\bm{\theta}; d_i)}{M_{total}} \\
&=& \prod\limits_{i=1}^{N} \frac{f(g, \Phi_0, M_{total}, r_h ; r_i,v_i)}{M_{total}},
\label{eq:likelihood}
\end{eqnarray}
where the stars are assumed to be independent.

For lowered-isothermal models, the DF $f$ is calculated numerically via the \texttt{limepy} software \citep{Gieles2015}, and thus the likelihood must be calculated numerically too.

As mentioned in Section~\ref{sec:data}, we simulate position and kinematic data of stars following a \texttt{limepy} model DF with parameters $\mathbf{\bm{\theta}}$ shown in Table~\ref{tab:partable}, assume the likelihood defined in equation~\ref{eq:likelihood}, and define physically-motivated informative priors on the model parameters.

Given that the likelihood is defined by the DF that was used to generate the data, we expect to obtain reasonable parameter estimates through inference made from the posterior distribution using Markov Chain Monte Carlo (MCMC) sampling. However, we are also going to impose prior distributions that are at least weakly informative, and so it is good practice to test whether the posterior can still be used to reliably infer the model parameters. Moreover, in the cases where the sampling of stars from the cluster is biased to inside the core or outside the core, we aim to understand how this sampling bias affects parameter inference.

\subsection{Prior Distributions}\label{sec:priors}

Two advantages of Bayesian inference are the necessity to incorporate meaningful prior information, and the requirement to state this explicitly. In order for the DF to correspond to a physically realistic collection of stars in a GC, all model parameters must be greater than zero. 
Negative parameter values are not allowed by the likelihood, but we also disallow negative parameter values via the priors (this increases efficiency and keeps the \texttt{limepy} model from returning errors).

One reason to use informative priors is that images and studies both within the Milky Way Galaxy and around other galaxies provide prior information on quantities like the mass and half-light radius of GCs. For example, GC masses span about an order of magnitude and most astronomers would be comfortable setting the prior $p(\log_{10}M_{total}) \sim N(\mu_M, \sigma_M)$, where the hyperparamters\footnote{the term hyperparameters is used to differentiate $\mu_M$ and $\sigma_M$ from the model parameters of interest} $\mu_M$ and $\sigma_M$ are defined in $\log_{10}M_{total}$. This is the prior we choose, and it is also supported by the near universal GC mass function \citep{Brodie06,Harris10}.

The \texttt{limepy} model works in $M_{total}$ space, so we need to do a change of variables to obtain the prior $p(M_{total})$. 
Using a change of variables, the prior on $M_{total}$ is
\begin{equation}
    p(M_{total}) = \frac{N(\mu_M,\sigma_M)}{M_{total}\ln10}.
\end{equation}

The half-light radius is another quantity of GCs for which we have considerable prior information. Images of GCs give an independent estimate of $r_h$, with a conservative measurement uncertainty of roughly 0.4pc \citep[e.g.][]{deBoer2019}. In this simulation study, we assume the observer has this prior information and set a truncated normal prior on $r_h$. 

We have considerably less prior information on the values of $g$ and $\Phi_0$, aside from the physically allowable, positive values. For these parameters, we use truncated uniform distributions. In summary, we assume the parameters for the \texttt{limepy} model are distributed as
\begin{eqnarray}
g & \sim & \text{unif}(0.001, 3.5), \\
\Phi_0 & \sim &\text{unif}(1.5, 14), \\
M_{total} & \sim & \frac{N(\mu_M, \sigma_M)}{M_{total}\ln(10)}, \\
\text{and  } r_h & \sim & N(a, b, \mu_{r_h}, \sigma_{r_h}),
\end{eqnarray}
where $\mu_M = 5.85$ and $\sigma_M=0.6$ (defined in $\log_{10}M_{total}$), and hyperparameters for the lower and upper bounds of $r_h$ are $a=0$ and $b=30$ respectively. The mean and standard deviation for the $r_h$ parameter ($\mu_{r_h}$ and $\sigma_{r_h}$) are chosen to reflect plausible information an observer would have for a given GC. Thus, for the average GCs in our analysis, we try different means, such as $\mu_{r_h}=3.4$, $\mu_{r_h}=3.1$, etc. with $\sigma_{r_h}=0.4$pc. Our results are insensitive to the choice of the mean, as long as it is not too many standard deviations away from the true value.

\subsection{Sampling the Target Distribution}\label{sec:mcmc}

Given the \texttt{limepy} model, we have a likelihood function $\mathcal{L}(g, \Phi_0, M, r_h \, ;\bm{d})$ for the four unknown parameters, depending on the observed star data $\bm{d}$. Combining the above prior distributions with this \texttt{limepy} likelihood function leads to a posterior distribution or \textit{target posterior density} via Bayes' theorem (eq.~\ref{eq:bayes}),
\begin{align*}
    p(g, \Phi_0, M_{total}, r_h | \bm{d}) &\propto p(\bm{d} | g, \Phi_0, M_{total}, r_h) \times \\ & p(g)p(\Phi_0)p(M_{total})p(r_h),
    \label{eq:bayes}
\end{align*}
where we assume independent priors.  Our goal is to sample from the target distribution $p(g, \Phi_0, M_{total}, r_h | \bm{d})$, and perform inference of the parameter values and the cumulative mass profile of the GC.

Ultimately, we explore and collect samples of this posterior density using a Markov Chain Monte Carlo (MCMC) algorithm, specifically a version of the standard Metropolis algorithm (Metropolis et al., 1953) that includes automated, finite adaptive tuning (to be discussed later). First, however, we find optimal starting values; we use the differential evolution optimizer function \texttt{DEopt} from the \texttt{NMOF} package \citep{NMOF,NMOFbook} in \textbf{R} \citep{Rbase} to find modal (i.e., argmax) values of the four parameters, and then use these values as the initial state of our MCMC algorithm. Differential evolution was first introduced by \cite{storn1997differential}, and we refer the reader to this paper for details on the algorithm. This initial step allows an automated selection of good starting values, which helps to overcome the complicated structure of the posterior distribution, thereby making sampling more efficient. Once the starting values are obtained, we run an automated, finite adaptive-tuning method during the burn-in of the Markov chain. To describe the finite adaptive-tuning method, we first provide a brief review of proposal distributions and sampling efficiency.

Sampling a target or posterior distribution using a standard Metropolis algorithm requires a choice of proposal or ``jumping'' distribution. The latter is used to randomly suggest a new place in parameter space, $\theta^*$, based on the current location $\theta_i$. Often, this suggestion is done using a normal distribution such that 
\begin{equation}
    \bm{\theta}^* = \bm{\theta}_i + \bm{Z},
\end{equation}
where $\bm{Z} \sim N(0,\bm{\Sigma})$. Here, $N(0,\bm{\Sigma})$ is the jumping distribution with a covariance matrix $\bm{\Sigma}$ set by the user. The value of $\Sigma$ determines whether, on average, ``big jumps'' or ``small jumps'' are attempted from the current location of $\theta_i$. These proposed jumps are either accepted or rejected according to the standard formula in the Metropolis algorithm. The efficiency of the sampling is dependent on the choice of this covariance matrix. For example, if the variance is too small then the algorithm make jumps that are too small. If the variance is too large, then the algorithm will make jumps that are too large. 

Finding a $\bm{\Sigma}$ that enables the most efficient sampling is sometimes accomplished through manual \textit{tuning}: adjusting $\bm{\Sigma}$ until the appropriate acceptance rate is achieved. Obviously, this can be a tedious and time-consuming process, especially in the case of multiple parameters. Thankfully, there are methods which automate this task and that are founded in statistical theory.

In this paper, we use an \textit{automated, finite adaptive-tuning} method during the burn-in of the Markov chain. This adaptive-tuning method is one in which the proposal step sizes are adjusted automatically and iteratively. We obtain a good covariance matrix for the proposal distribution using an Adaptive Metropolis algorithm (Haario et al., 2001; Roberts and Rosenthal, 2009) which repeatedly updates the Metropolis proposal distribution (i.e., the proposal covariance matrix) based on the empirical covariance of the run so far, in an effort to obtain a proposal covariance matrix equal to about $(2.38)^2$ times the the target covariance matrix divided by the Markov chain's dimension, which has been shown to be optimal under appropriate assumptions \citep{roberts1997geometric, roberts2001optimal}. Foundational works on the subject of adaptive Metropolis and convergence are found in the statistics literature \citep{haario2001adaptive, roberts1997weak, roberts2009}.

The practice of using the Adaptive Metropolis algorithm for an \textit{initial} run and then fixing the proposal variance for the final run corresponds to ``finite adaptation'' as in Proposition~3 of \cite{roberts2007coupling}. We require a minimum of five initial runs to update the proposal variance, but also automatically allow for further iterations as needed to achieve efficient sampling. Almost all of the GCs we analyze take no more than five iterations of the finite adaptive tuning, which takes one to five minutes per cluster on a simple laptop computer.

Once the finite adaptive step is complete, we run a standard Metropolis algorithm using the final (hopefully approximately optimal) proposal distribution found by the Adaptive Metropolis step. The final sampling takes less than 15 minutes per cluster to complete. At the end, we discard an initial burn-in period, and take the remaining chain values as a sample from the posterior density.

The above procedure allows us to approximately sample from $p(g, \Phi_0, M_{total}, r_h | \bm{d})$, and hence (a) approximately compute the posterior means and other statistics of the four unknown parameters $(g, \Phi_0, M_{total}, r_h)$, including Bayesian credible intervals, and (b) calculate a cumulative mass profile of the GC for every sample from the target distribution.

\subsection{Different Cluster and Sampling Cases}

Very generally, GCs may be classified as having an average, compact, or extended morphology based on their radius $r_h$. Additionally, the spatial and kinematic data from stars may be a random sample from everywhere in the cluster, a random sample beyond some radius, or a random sample within some radius. We expect the ability of our method to recover the true mass and mass profile to depend on both GC morphology and the type of sampling of its stars. Understanding the bias in parameter inference that can occur as a result of biased sampling is important, since in reality we sometimes lack position and kinematic data from the inner or outer regions of the cluster. Thus, we investigate multiple combinations of the aforementioned cases to understand any possible bias. 

\begin{table}[]
    \centering
    \begin{tabular}{l|c|c|c}
        \sc{GC Type} & \multicolumn{3}{c}{\sc{Sampling}}\\
        & \textbf{random} & \textbf{outer regions} & \textbf{inner regions}\\
        \hline
        average & \checkmark & \checkmark & \checkmark \\
        compact & \checkmark  & \checkmark & \checkmark \\
        extended & \checkmark & \checkmark & \checkmark \\
   high $\Phi_0$ GC & \checkmark & \checkmark & \checkmark \\
       low $\Phi_0$ GC & \checkmark & \checkmark & \checkmark \\
      \hline
    \end{tabular}
    \caption{Summary of analyses completed in this study.}
    \label{tab:cases}
\end{table}

Table~\ref{tab:cases} summarizes the combinations we investigate. For each case, we simulate 50 GCs using the parameter values listed in Table~\ref{tab:partable}, and subsample 500 stars either (1) randomly, (2) outside the $r_{cut}$ value, or (3) inside the $r_{cut}$ value. For simplicity, we use $r_{cut}=1.5$pc, and assume that all GCs are at the same distance. We choose this radius as a cut-off mostly for simplicity and partly because recent work by the HSTPROMO Team indicates that proper motions are most often available for stars within the half-mass radius \citep{Watkins2013} but not beyond. Our conservative choice for $r_{cut}$ is therefore half of the average effective radius of Galactic clusters (excluding very extended clusters with effective radii greater than 10 pc) \citep{Baumgardt2018}. In our simulated GCs, all stars have the same brightness and mass, and so the half-light radius corresponds to the half-mass radius.

By repeating the analysis on 50 randomly generated GCs, we estimate and examine the coverage probabilities for the Bayesian credible regions in the different scenarios listed in Table~\ref{tab:cases} (Section~\ref{sec:results}).

For example, for the average cluster, we generate 50 simulated GCs with parameter values $g=1.5, \Phi_0=5, M_{total}=10^5M_{\odot}$, and $r_h=3.0$pc, and randomly sample 500 stars from each GC. For each GC, we run the analysis on the subsample of stars, obtaining samples of the target distribution as described in the previous section. Next, we estimate the mean, interquartile range, and 95\% credible interval of the posterior distribution using our MCMC samples from the target distribution. After doing this for all 50 average GCs, we count how many times the interquartile ranges and 95\% credible intervals cover the true parameter value to estimate the coverage probability. If the Bayesian credible regions are reliable, then the interquartile ranges should cover the true parameter values 50\% of the time, and the 95\% credible intervals should cover the true parameter values 95\% of the time.

The same procedure is repeated for GCs with different half-light radii, reflecting extended and compact clusters. For these clusters we use parameter values of $g=1.5, \Phi_0=5, M_{total}=10^5M_{\odot}$, and $r_h=9.0$pc and $g=1.5, \Phi_0=5, M_{total}=10^5M_{\odot}$, and $r_h=1.0$pc respectively (i.e. the scenarios listed in Table~\ref{tab:cases}). To further explore the parameter space believed to be covered by Galactic GCs, and specifically to explore GCs that are more (less) concentrated, we also look GCs with a high (low) $\Phi_0$.

Using our estimate of the posterior distribution for a single GC, we can also estimate that GC's cumulative mass profile (CMP). The CMP is an estimate of the mass contained within some distance $r$ of the GC. To estimate the CMP, we follow the same procedure as described in \cite{eadie2019cumulative}, who used this approach to estimate the Milky Way's CMP. For every set of model parameters $(g, \Phi_0, M_{total}, r_h)$ sampled by our algorithm (i.e., every row of parameter values in the Markov chain), we calculate the cumulative mass profile determined by the \texttt{limepy} model. Because we have 1000s of rows in our Markov chain, we obtain thousands of CMP estimates. These CMPs provide us with a visual and quantitative estimate that can be used to calculate Bayesian credible regions and that can be compared directly to the true CMP of the cluster.

In all of the above examples, we have assumed that we know the complete position and velocity components of the stars. However, in reality we often have incomplete data. For example, we may only have projected measurements on the plane of the sky (i.e., projected distances in the $x-y$ plane, and proper motions). This missing data may influence our mass and mass profile estimates in unexpected ways, and is important to study. In a Bayesian analysis one can treat the missing components as parameters in the model, but this also means that further prior distributions must be set. Given the complexity of the problem, we leave this to future work. 

\section{Results \& Discussion}\label{sec:results}

\subsection{Random Sampling}

For the cases in which we randomly sample stars from everywhere in the cluster (i.e., the first column in Table~\ref{tab:cases}), we find the Bayesian credible regions to be reliable for the average, extended, and compact GCs. 

\begin{figure*}
    \centering
    \includegraphics[width=1\textwidth, trim=0cm 1cm 0cm 0cm, clip]{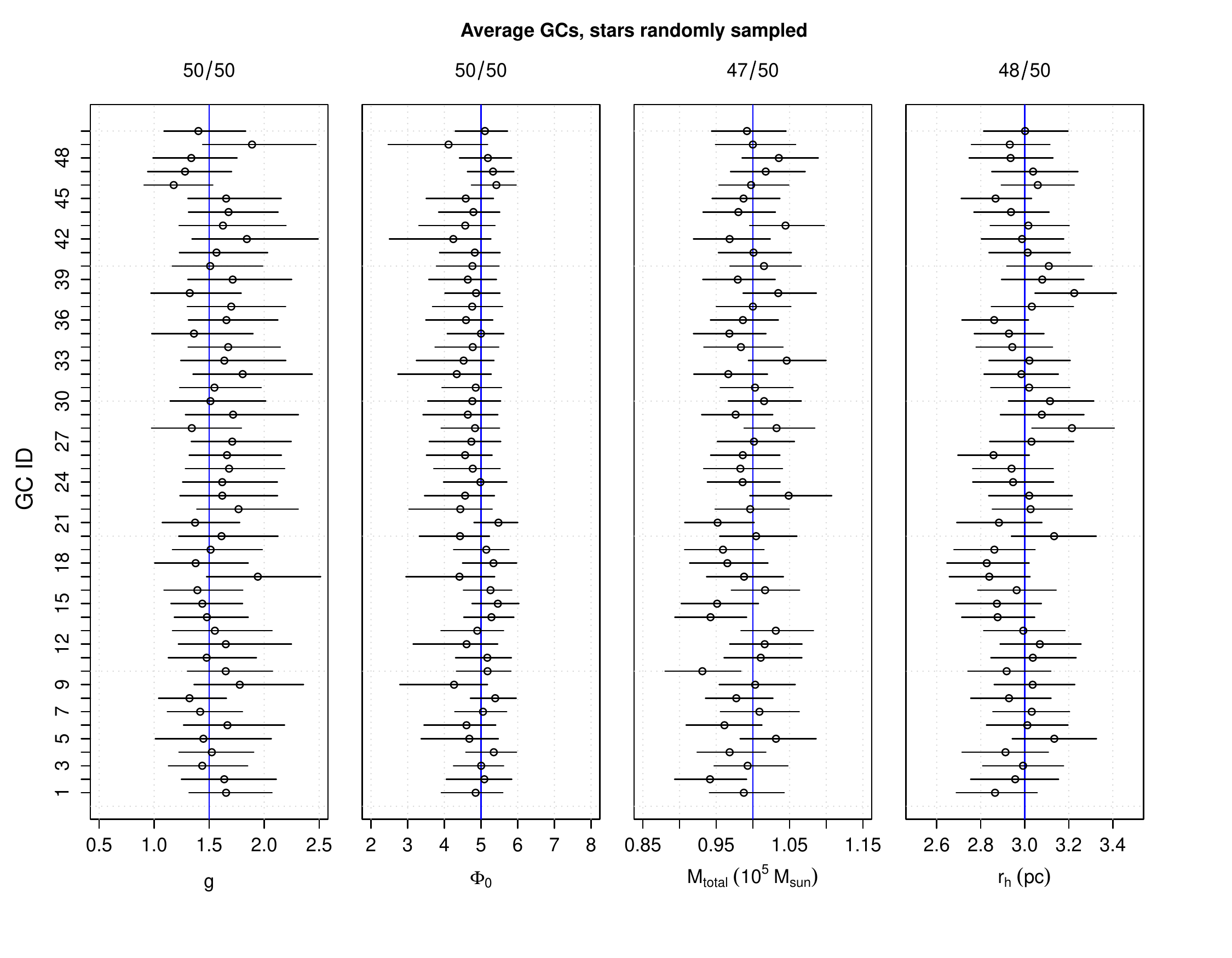}
    \caption{The parameter estimates for fifty simulated ``average'' GCs, and their 95\% credible intervals. Each panel shows 50 credible intervals (error bars), the corresponding mean (points), and the true parameter value (vertical blue line). Each \textit{row} of points across the four panels corresponds to the parameter estimates for the GC with ID given on the vertical axis. The fraction at the top of each panel indicates the number of times the 95\% credible interval overlaps the true parameter value. The fractions are very large, as they should be for 95\% intervals.}
    \label{fig:95avg}
\end{figure*}

As an example, Figure~\ref{fig:95avg} shows the 95\% credible intervals  (error bars) for each model parameter, for 50 realizations of an average cluster. The true parameter values are shown as vertical blue lines, and the number of times out of 50 that the 95\% credible interval of the target distribution overlaps the true value is shown at the top of each panel. We can see that the credible intervals for each parameter reliably contains the true parameter approximately 95\% of the time (Figure~\ref{fig:95avg}).

\begin{table}[]
    \centering
    
    \begin{tabular}{l|c|c}
     \sc{GC Type} & \sc{C.I.} & \sc{Coverage Prob. for $M_{total}$} \\
          \hline
        average &  & 0.50 \\
        compact&  & 0.42  \\
        extended & 50\% & 0.52  \\
        high $\Phi_0$ & & 0.48 \\
        low $\Phi_0$ & & 0.38 \\
        \hline
        \hline
        average &  & 0.94  \\
        compact&  & 0.90  \\
        extended & 95\% & 1.00  \\
        high $\Phi_0$ & & 0.94 \\
        low $\Phi_0$ & & 0.92 \\
    \end{tabular}
    \caption{\textit{Estimated coverage probabilities under the random sampling case, for different GC morphologies.}}.
    \label{tab:MassBias}
\end{table}

\begin{figure*}
    \centering
    \includegraphics[width=1\textwidth, trim=0cm 1cm 0cm 0cm, clip]{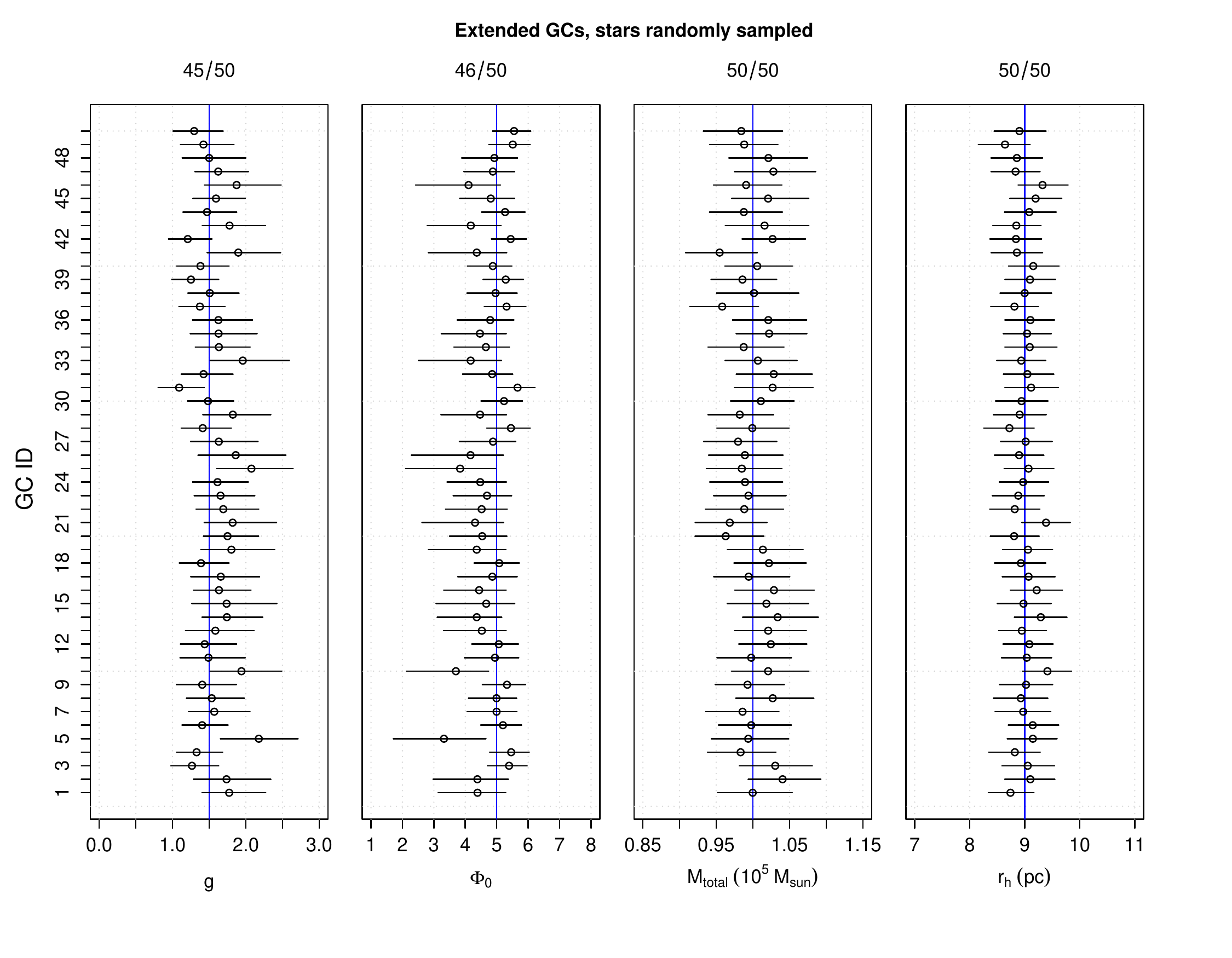}
    \caption{Same as Figure~\ref{fig:95avg}, but the 95\% credible intervals for the ``extended'' GCs whose stars are randomly sampled at all radii. The fractions are again very large, as they should be.}
    \label{fig:50quants_extended}
\end{figure*}

As a second example, we show a similar plot for the case of the extended GCs (Figure~\ref{fig:50quants_extended}). Here too, we find the 95\% credible intervals to be reliable for the most part. The credible intervals for $g$ and $\Phi_0$ are slightly overconfident, since the true parameter value lies within the 95\% credible intervals only 90\% and 92\% of the time respectively.

\begin{figure*}
    \centering
    \includegraphics[width=1\textwidth, trim=0cm 1cm 0cm 0cm, clip]{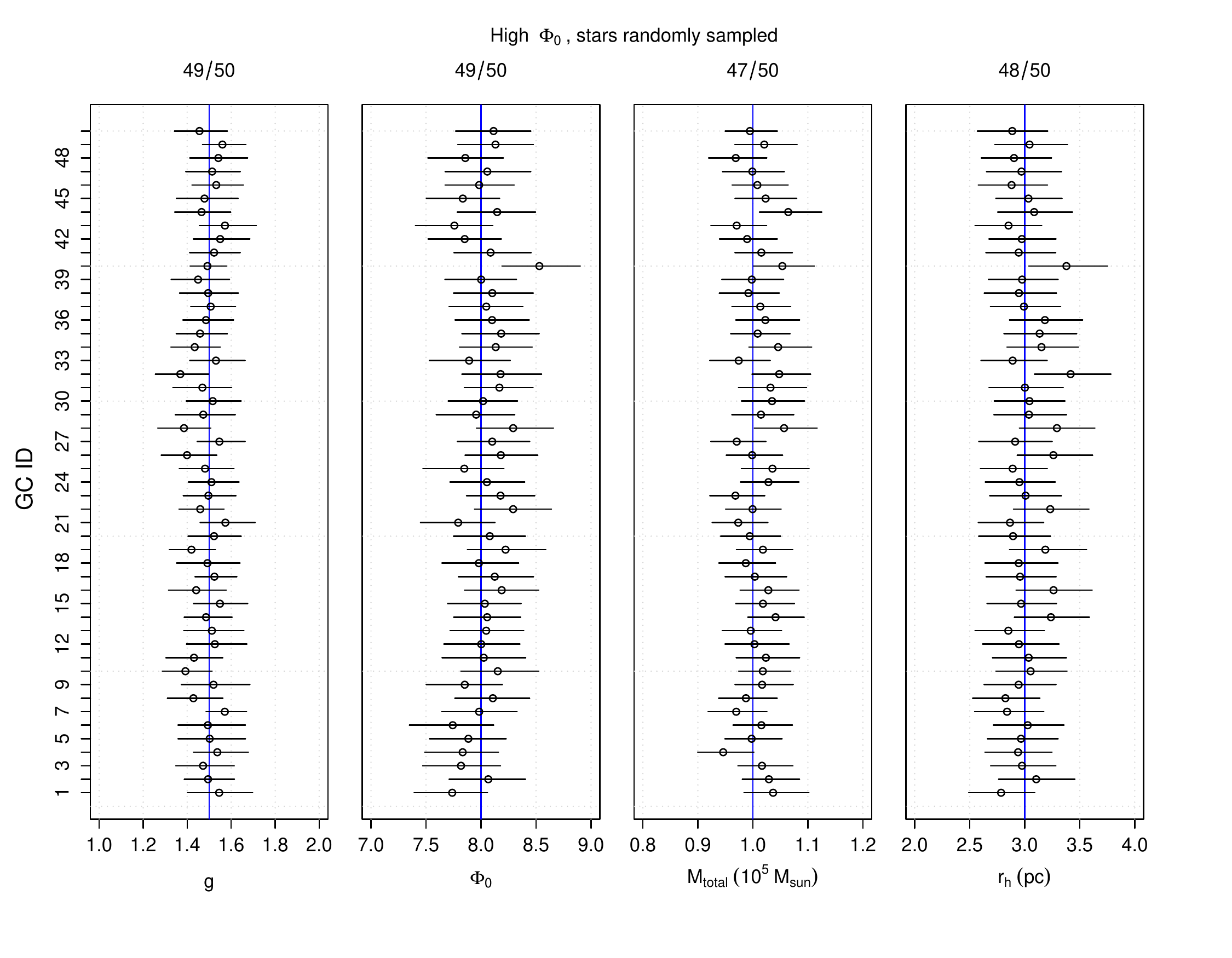}
    \caption{Same as Figure~\ref{fig:95avg}, but the 95\% credible intervals for the GCs with a high $\Phi_0$, when stars are randomly sampled at all radii. The fractions are again very large, as they should be.}
    \label{fig:95quants_highPhi0}
\end{figure*}

As a final and third example, Figure~\ref{fig:95quants_highPhi0} shows the same type of plot for a more concentrated cluster with $\Phi_0=8$. Again, the credible intervals are reliable, showing good coverage probabilities.

Table~\ref{tab:MassBias} shows the estimated coverage probabilities for the $M_{total}$ parameter in the case of random sampling, for all three types of clusters, found by calculating the fraction of times that the true $M_{total}$ is contained within the Bayesian credible interval. We can see that both the 50\% and 95\% credible intervals for $M_{total}$ are reliable when the stars are randomly sampled throughout the cluster, despite cluster type.

\begin{figure*}
    \centering
    \includegraphics[width=0.9\textwidth, trim=0cm 0.5cm 0cm 0cm, clip]{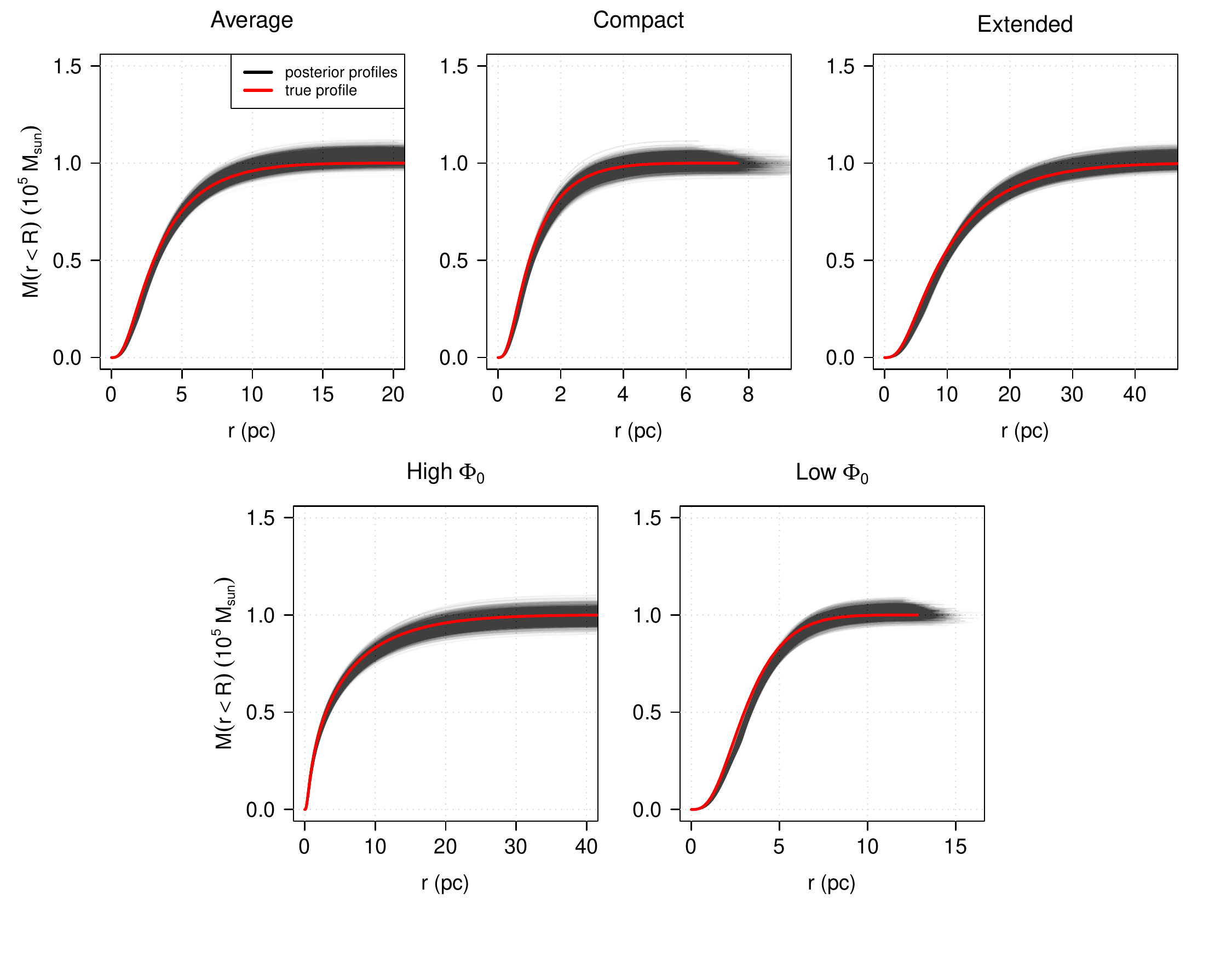}
    \caption{Example mass profiles calculated from the posterior samples (black curves) for the cases of random sampling listed in Table~\ref{tab:cases}. The type of GC (average, compact, extended, high $\Phi_0$, or low $\Phi_0$) are shown above the figure. Each plot shows the posterior samples for a single GC. The red solid curves show the true mass profile from the \texttt{limepy} model, showing excellent agreement.}
    \label{fig:massprofilesrandom}
\end{figure*}

The MCMC samples can also be used to infer the cumulative mass profile (CMP) of the cluster under the \texttt{limepy} model. Figure~\ref{fig:massprofilesrandom} shows the CMP inferred for each of an average, compact, extended, low $\Phi_0$, and high $\Phi_0$ cluster in the random sampling case. The posterior distribution of $g, \Phi_0, M_{total}$ and $r_h$ from the Markov chains are used to calculate the posterior estimate of the CMP, shown as transparent black curves. The red curve shows the true CMP given by the \texttt{limepy} model with the correct parameters. We can see that in all cases, the CMP is recovered quite well.

In general, we find that the the 50\% and 95\% credible regions and CMPs are reliable for all types of GCs when the stars are sampled randomly throughout the cluster. It is reassuring that we can recover the true parameter values and the CMPs reliably from a random sample of only 500 stars.

\begin{figure*}
    \centering
    \includegraphics[width=0.99\textwidth, trim=0cm 0.75cm 0cm 0cm, clip]{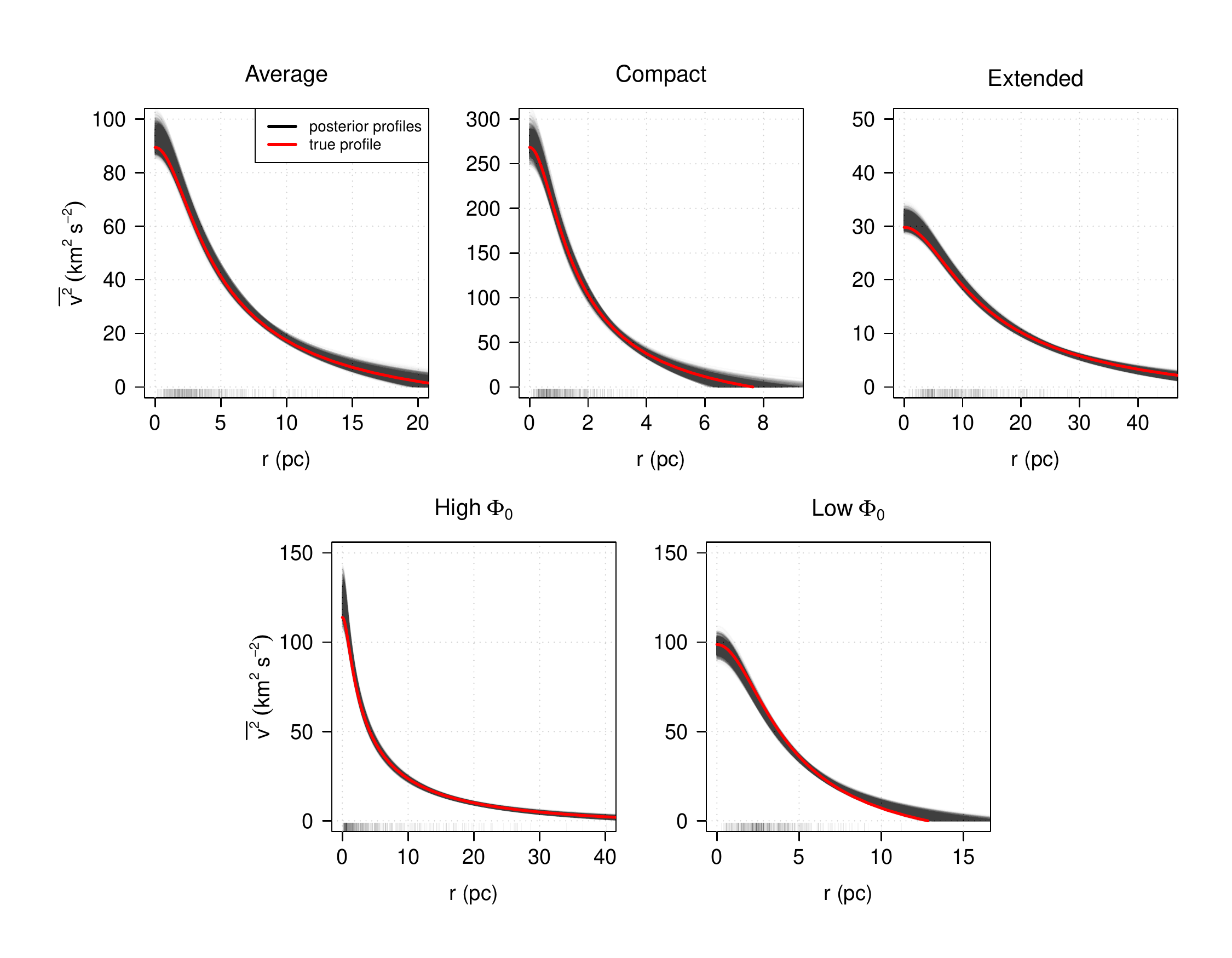}
    \caption{Example mean-square velocity profiles calculated from the posterior samples (black curves) for the case of random sampling. Each plot title indicates the type of GC (average, compact, extended, high $\Phi_0$, or low $\Phi_0$). The red solid curves show the true mean-square velocity profile from the \texttt{limepy} model, again showing good agreement. The semi-transparent vertical dashes along the bottom of each plot show the exact location $r$ of the randomly sampled stars in the GC.}
    \label{fig:velocity_profile}
\end{figure*}

Additionally, we can inspect other physical quantities provided by the \texttt{limepy} model fit. For example, Figure~\ref{fig:velocity_profile} shows the mean-square velocity $\overline{v^2}$ profiles as a function of radius for one GC in each of the five morphologies. Under random sampling of the stars, we observe that the true mean-square velocity profile is well-recovered by the MCMC samples.

\subsection{Biased Sampling}

In general, we find that biased sampling of stars from only inside or outside the cluster core results in model parameter estimates that are biased and in Bayesian credible intervals that are unreliable. While obtaining biased estimates from a biased data sample is not surprising, the reality is that this type of sampling mimics the data from some telescopes. Investigating these cases can illuminate the kind of biases we should expect and possibly correct for. Indeed, through our investigations of biased sampling, we find the success of the parameter inference and CMP inference is a combination of \textit{both} the cluster's morphology \textit{and} the type of biased sampling. 

\begin{figure*}
    \centering
    \includegraphics[width=1\textwidth, trim=0cm 1cm 0cm 0cm, clip]{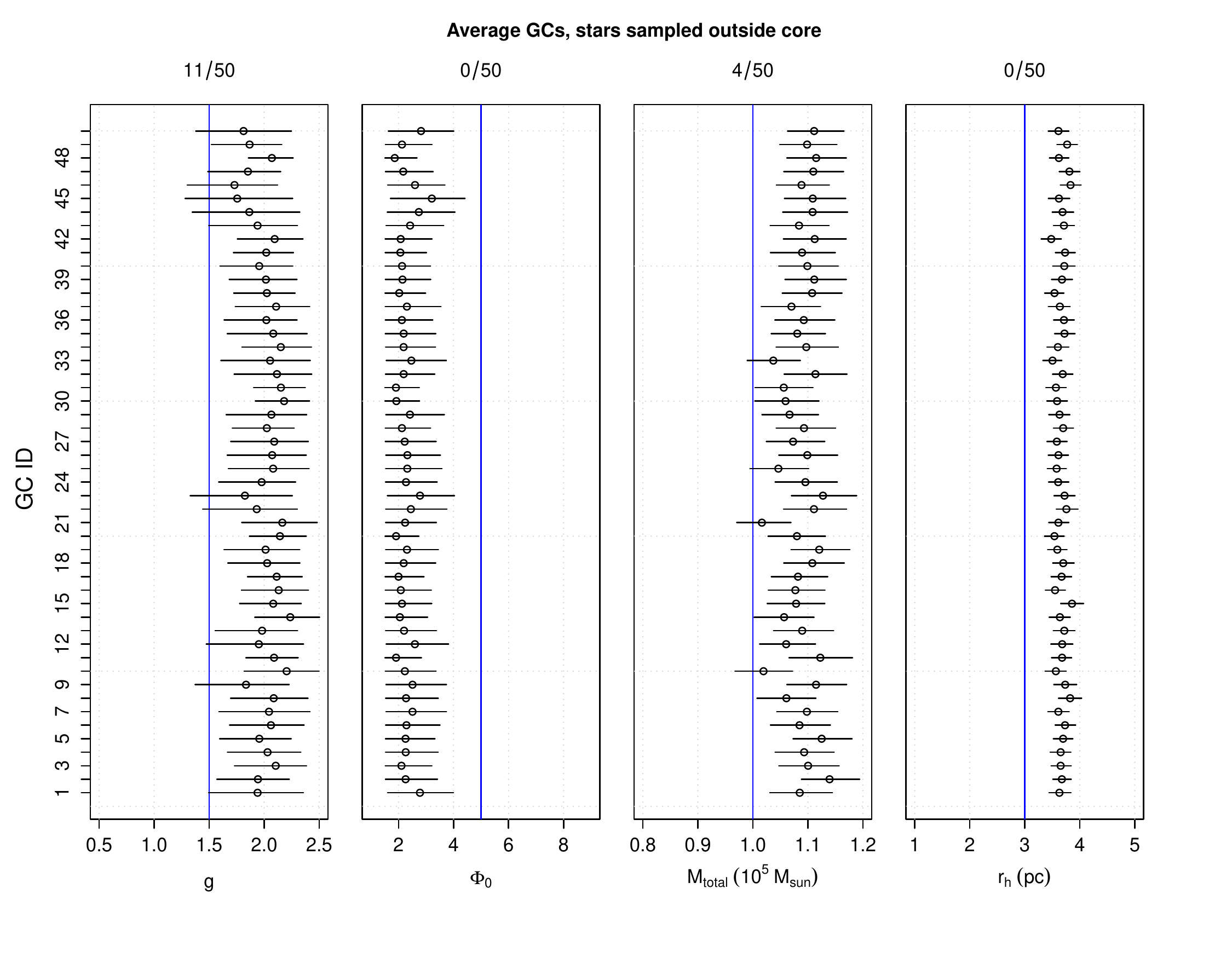}
    \caption{The mean estimate and 95\% credible intervals for average GCs whose stars were sampled outside the core. Due to the biased sampling, most of the intervals miss the true value.}
    \label{fig:avgGCbiased}
\end{figure*}

As an example, Figure~\ref{fig:avgGCbiased} shows the 95\% credible intervals for an average GC when only the outer stars' data are sampled. We can see that the interquartile ranges are unreliable, and that parameter estimates are biased. In particular, $M_{total}$, $g$, and $r_h$  are consistently overestimated, while $\Phi_0$ is underestimated.

\begin{figure*}
    \centering
    \includegraphics[width=1\textwidth, trim=0cm 1cm 0cm 0cm, clip]{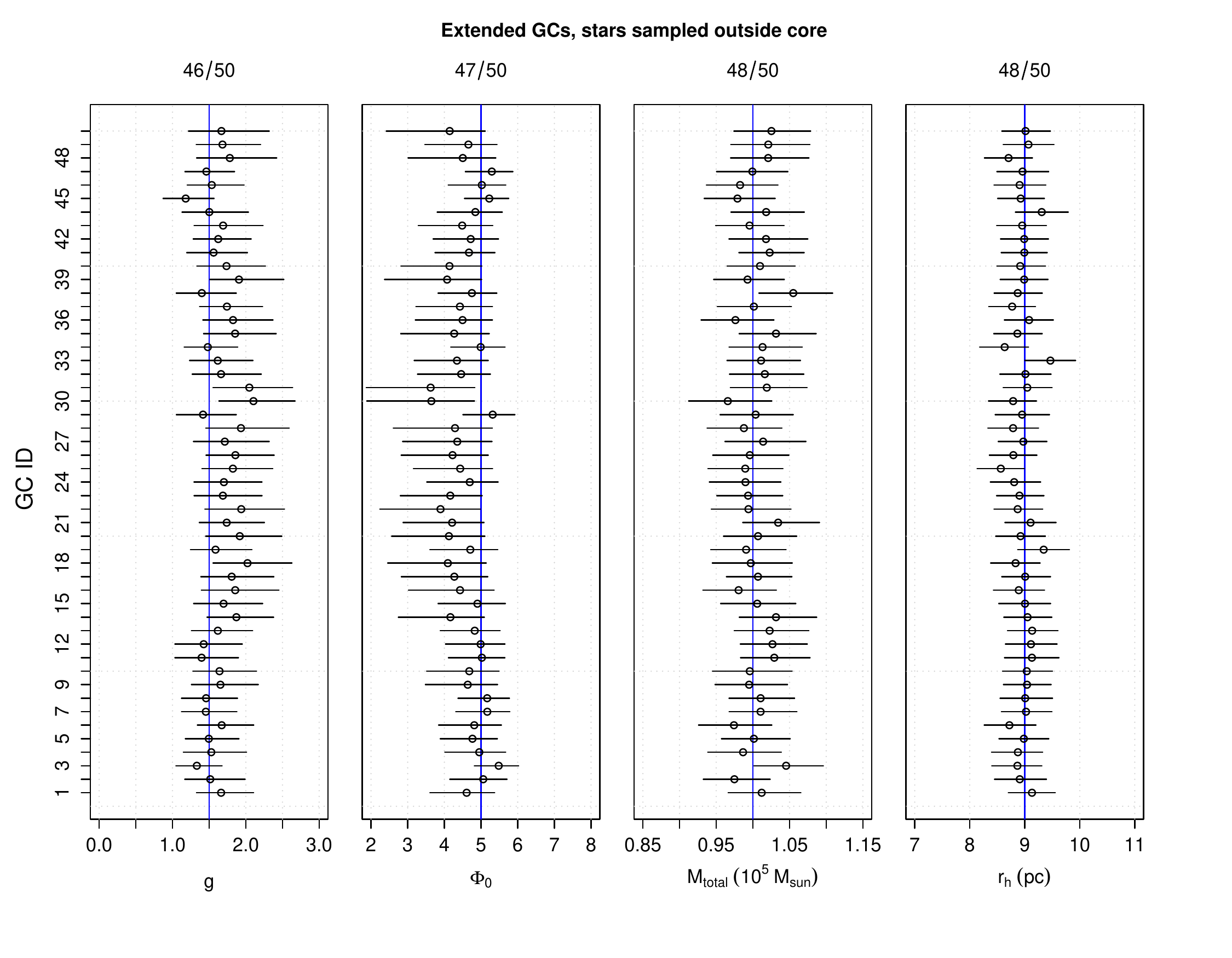}
    \caption{Same as Figure~\ref{fig:95avg}, but for extended GCs whose stars are sampled only from the outer regions.}
    \label{fig:extended_outer}
\end{figure*}

In contrast, biased sampling of outer stars of an extended cluster result in parameter estimates that are much more reliable (Figure~\ref{fig:extended_outer}). In this case, the extended GC's mass $M_{total}$ and half-light radius $r_h$ can actually be estimated reliably.

In Table~\ref{tab:MassBias_biased}, we summarize how reliably we can recover $M_{total}$ in the biased sampling cases. Only the extended cluster with sampling in the outer regions is reliable. Also note the $*$ in the table, which indicate cases in which our MCMC algorithm has trouble finding a stationary distribution with good mixing, leading to biased estimates of the total mass (Table~\ref{tab:MassBias}). In these particular cases, the behaviour of the Markov chain would be a clue to the observer that the model is having trouble describing the data.

\begin{table}[]
    \centering
    \begin{tabular}{l|c|c|c}
     \sc{GC Type} & \sc{C.I.} & \multicolumn{2}{c}{\sc{Coverage Prob. for $M_{total}$}} \\
           & & \textit{\small{outside core}} & \textit{\small{inside core}} \\
        \hline
        average &  & 0.02*, +  & 0.00*, $-$ \\
        compact& & 0.00*, + & 0.14*, $-$ \\
        extended & 50\% & 0.60, $-$ & 0.00*, $-$ \\
        high $\Phi_0$ & & 0.00, + & 0.00, $-$\\
        low $\Phi_0$ & & 0.12, + & 0.00*, $-$ \\
        \hline
        \hline
        average &  & 0.08*, +  & 0.00*, $-$ \\
        compact & & 0.00*, + & 0.62*, $-$ \\
        extended &  95\% & 0.96, $-$ & 0.00*, $-$ \\
       high $\Phi_0$ & & 0.00, + & 0.00*, $-$ \\
        low $\Phi_0$ & & 0.48, + & 0.00*, $-$ \\
    \end{tabular}
    \caption{\textit{Estimated coverage probabilities and bias in mass estimates.} Also shown is whether the mass parameters are on average overestimated ($+$) or underestimated ($-$), or unbiased (no symbol). A $*$ indicates the chains had trouble converging and/or the estimates are at the lower or upper end of the prior distribution(s).}
    \label{tab:MassBias_biased}
\end{table}

\begin{figure*}
    \centering
    \includegraphics[width=\textwidth, trim=0cm 1cm 0cm 0cm, clip]{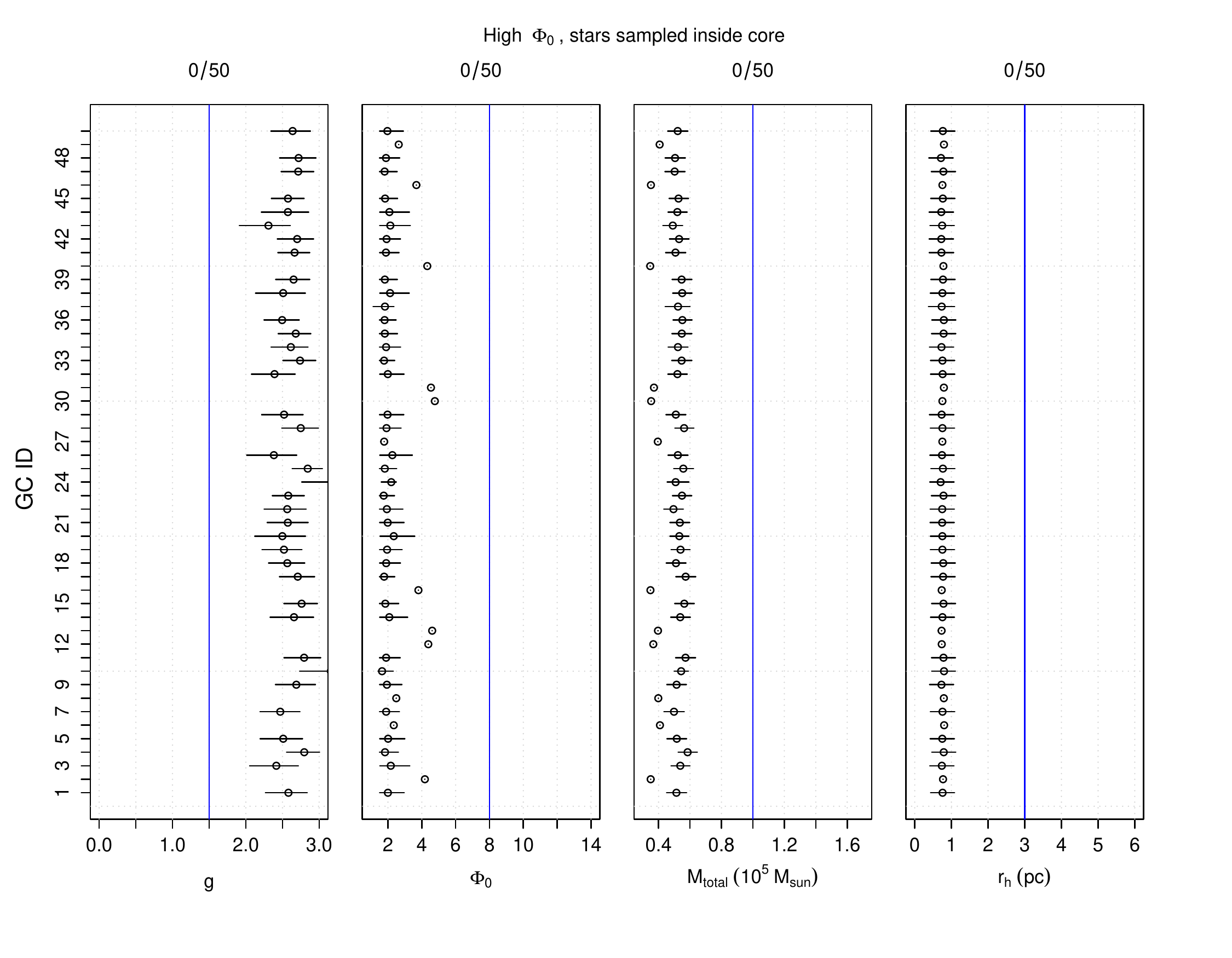}
    \caption{Same as Figure~\ref{fig:95avg}, but for high $\Phi_0$ GCs sampled inside the core.}
    \label{fig:95quants_highPhi0Biased}
\end{figure*}

For the concentrated GCs (i.e., high $\Phi_0$), biased sampling of stars in the inner regions also leads to poor parameter estimates and unreliable credible regions (Figure~\ref{fig:95quants_highPhi0Biased}). For some of the GCs in the scenario, the Markov chains become stuck in one location. The estimates of the mean from these bad chains are shown as the open circles with a small dot in the middle (i.e., the ``estimates'' have a variance of zero because the chains became stuck at a single place in parameter space). The exact estimated parameter values in these bad cases are rather meaningless and random. Moreover, if a scientist were to see this behaviour in a Markov chain from a real data analysis, then they would know not to trust the solution. However, in many cases of randomly generated GCs with high $\Phi_0$ and biased sampling in the inner regions, the Markov chains do look reasonable even when their estimates are not. Thus, a scientist could mistakenly assume the convergence is giving reliable parameter estimates. We will return to this scenario shortly.

\begin{figure*}
    \centering
    \includegraphics[width=0.9\textwidth, trim=0cm 0.5cm 0cm 0cm, clip]{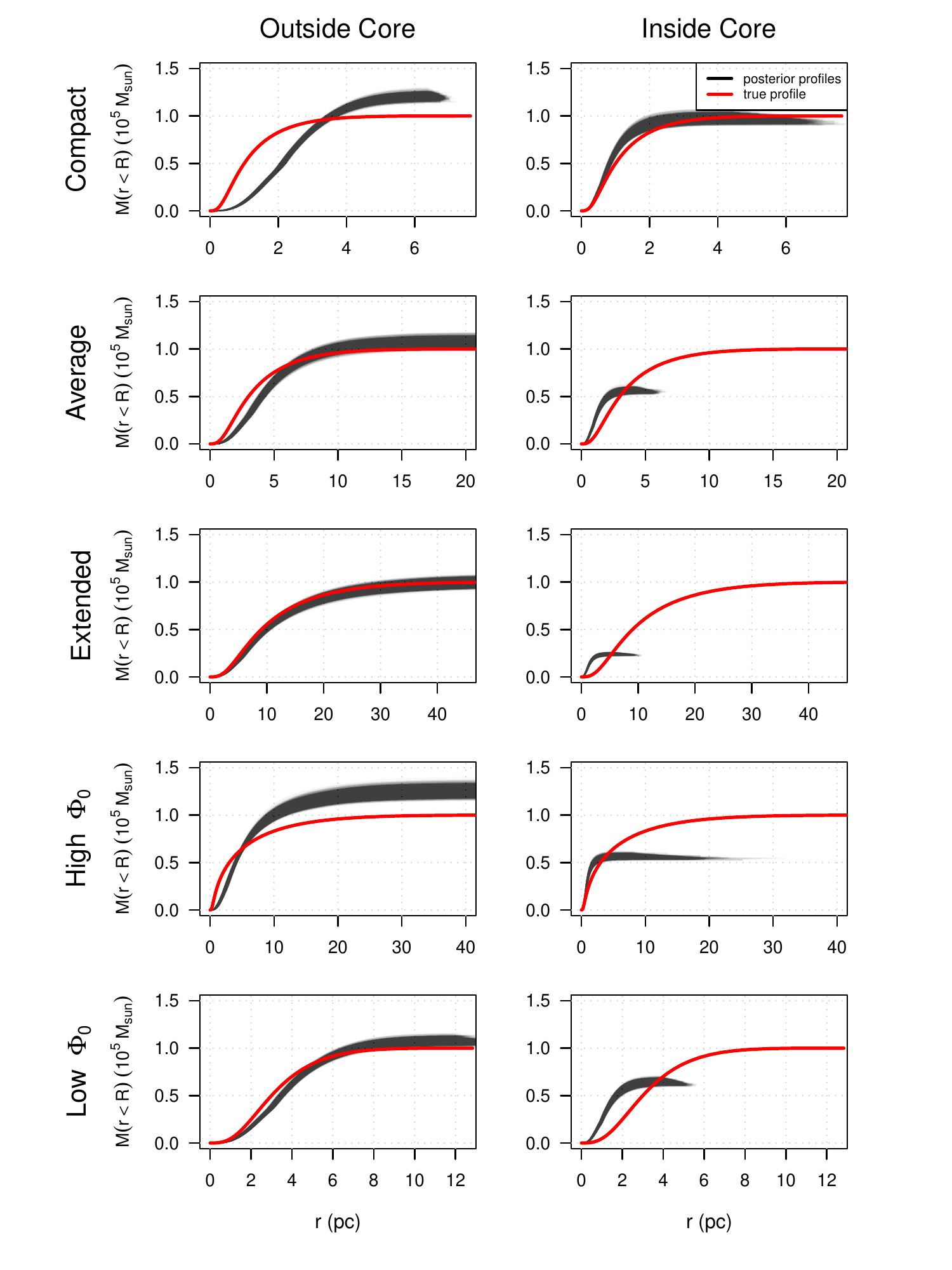}
    \caption{Example cumulative mass profile estimates (black curve) when stars are subject to selection bias either outside or inside the core of the GC. The black semi-transparent curves show the mass profiles predicted by the MCMC samples, and the solid red curves show the true mass profiles. Each row corresponds to the type of GC, and each column corresponds to the type of biased sampling --- stars sampled outside or inside the core of the GC. The biased samples lead to very poor estimates in most cases, with the exception of the morphology-sampling combinations of extended cluster-outside core, the compact cluster-inside core, and low-$\Phi_0$-inside core.}
    \label{fig:CMPbiasedsampling}
\end{figure*}

The CMPs provide more insight than simply looking at the parameter estimates and their credible intervals. Figure~\ref{fig:CMPbiasedsampling} shows example CMPs for all GC morphologies when the stars in these GCs are sampled only in their outer or inner regions (first and second column respectively). Looking at the first column in Figure~\ref{fig:CMPbiasedsampling}, we see that when stars are sampled \textit{outside} the core, the \textit{inner region} of the cluster's profile tends to be underestimated --- regardless of the GC morphology. The opposite is true for sampling \textit{inside} the core (the second column). At the same time, sampling outside the core tends to lead to an overestimate of the total mass, while sampling inside the core leads to a (sometimes severe) underestimate.

There are two exceptions to the observation that biased samples lead to biased CMPs, namely (1) when extended and low $\Phi_0$ clusters are sampled in the outer regions, and (2) when compact clusters are sampled in the inner regions. For the extended and low $\Phi_0$ GC, our method is able to recover the true CMP reasonable well when stars outside the core are sampled, whereas this is certainly not the case when stars inside the core are sampled. For the compact GC, we see the opposite case --- the CMP is reasonably-well estimated when the sample contains stars inside the core versus outside the core. 

These cases where biased samples still lead to unbiased samples are not surprising --- sampling stars in the outer region of an \textit{extended} or \textit{less concentrated} cluster will provide a better representation of the true stellar distribution than sampling stars in its core, because these types of GCs are less dense in their inner regions (Figure~\ref{fig:3GCs}). Likewise, sampling stars in the inner region of a \textit{compact} cluster will be a better representation of the true stellar distribution than a sample from the outer region because compact GCs are more dense towards their centers.

\begin{figure*}
    \centering
    \includegraphics[width=0.9\textwidth, trim=0cm 0.5cm 0cm 0cm,  clip]{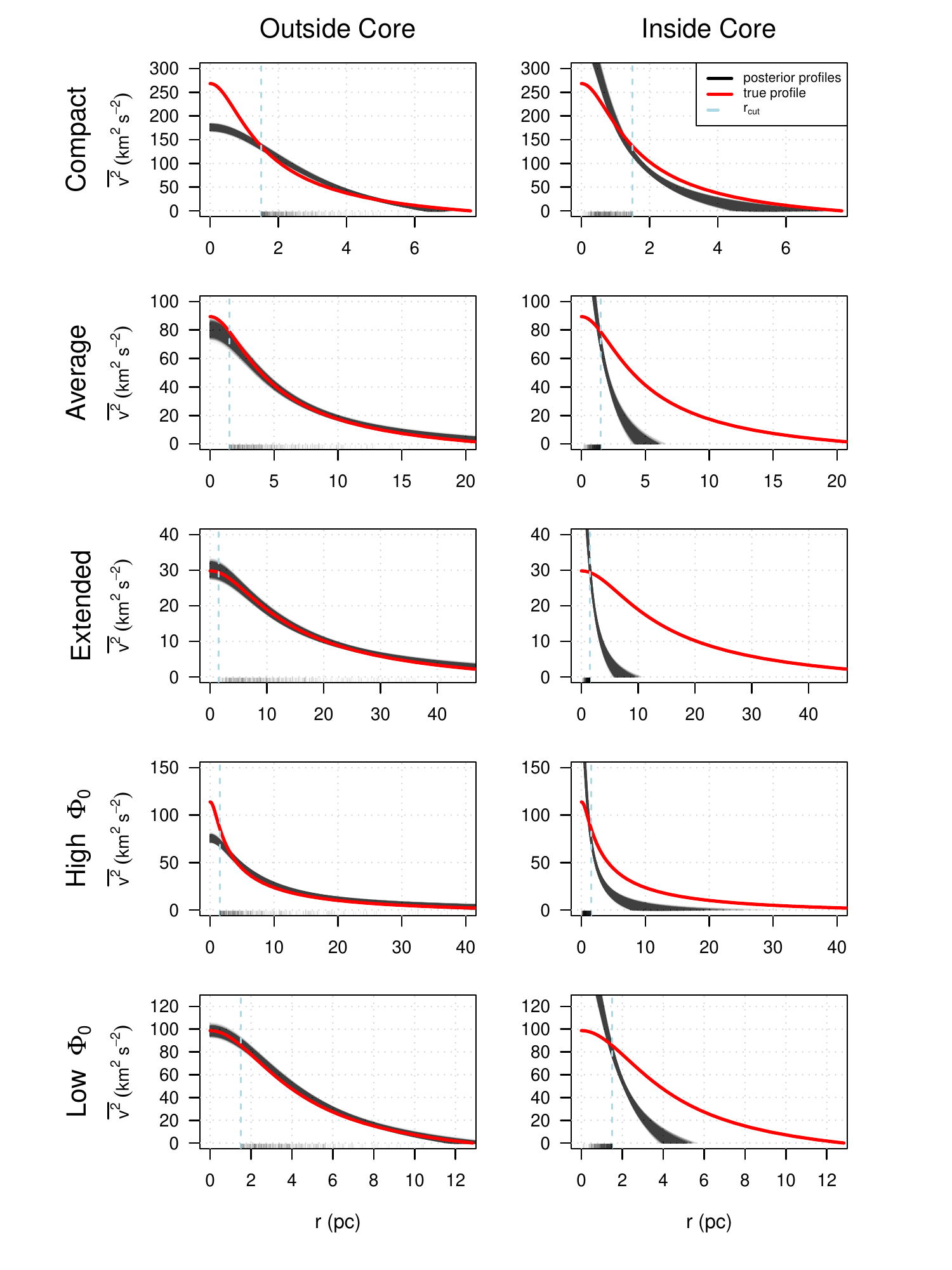}
    \caption{Example mean-square velocity profile estimates (black curves) from the MCMC samples when stars are subject to selection bias. The solid red curves show the true $\overline{v^2}$ profiles. Each row corresponds to the type of GC, and each column corresponds to the type of biased sampling --- stars sampled outside or inside the core of the GC. The vertical, light-blue dashed line indicates the $r_{cut}=1.5$pc and the semi-transparent vertical dashes along the bottom of each plot shows the individual positions of each star in the (biased) sample. The biased samples lead to very poor estimates in most cases, with the exception of the morphology-sampling combinations of average-outside core, extended cluster-outside core, and low-$\Phi_0$-outside core.}
    \label{fig:velprofilebiasedsampling}
\end{figure*}

Next, we use the MCMC samples to estimate the mean-square velocity $\overline{v^2}$.  In Figure~\ref{fig:velprofilebiasedsampling}, each row corresponds to a specific GC type, and the columns indicate whether stars were sampled outside (left) or inside (right) the core of the GC. The light blue, dashed line shows the $r_{cut}$ value, and along the bottom are semi-transparent marks showing the exact positions of the stars in the sample.

In the left-hand column of Figure~\ref{fig:velprofilebiasedsampling}, the estimated $\overline{v^2}$ profiles are reasonably-well matched to the true profiles for three morphologies (average, extended, and low-$\Phi_0$ GCs). Notably, the corresponding mass profile CMPs in Figure~\ref{fig:CMPbiasedsampling} are also some of better estimates of the entire set. For the other two types of GCs, it is the inner part of the profiles that do not match; the true mean-square velocity profile (red curve) in the center of the GC is much higher than the predicted profiles (black curves). Our findings suggest that a reasonable estimate of the $\overline{v^2}$ profile \textit{might} be possible for outer regions of the GC when stars are sampled outside the core, but that it would be ill-advised to extrapolate the model fit to the inner regions when only stars outside of the core are available.

In the right-hand column of Figure~\ref{fig:velprofilebiasedsampling}, we see that for every type of GC the true mean-square velocity profile is poorly matched by the predictions at all radii $r$. Within the $r_{cut}$ value, the true profile is generally lower than the black curves, whereas it is much higher than the black curves outside $r_{cut}$. Thus, the kinematic information from inner GC stars alone is not enough to constrain the model at any radii.

One aspect that we have not explored in the biased sampling cases is whether the $r_{cut}$ value plays a significant role in determining parameter estimates --- especially if that $r_{cut}$ value was more directly linked to GC morphology. Here, we have used a fixed $r_{cut}$ value mostly for simplicity --- but in future work it would be worth exploring the impact of $r_{cut}$ more fully. For example, the $r_{cut}$ value for an extended or less-concentrated GC might be relatively smaller than that for the $r_{cut}$ value for a compact or highly-concentrated GC.

It is also worth mentioning that for the fits in the right-hand column, the Markov chains had trouble converging and/or the estimates of the parameter were at the lower or upper end of the prior distributions (see Table~\ref{tab:MassBias_biased}). Both of these issues are red flags; the model has not been fit well to the data and any inference would be imprudent.

\section{Conclusion}\label{sec:conclusion}

This paper has investigated the estimation of globular cluster properties based upon a sample of their constituent stars.  We have developed a Markov chain Monte Carlo (MCMC) algorithm to compute the four parameters of a lowered isothermal model that is used to represent a GC system.  Our algorithm uses a version of the Metropolis algorithm, together with a numerical optimisation to find good starting values, and a finite adaptation tuning phase to find a good proposal covariance matrix.  We then applied our algorithm to simulated data generated using the \texttt{limepy} package \citep{Gieles2015}, and examined the extent to which the parameters, mass profile, and mean-square velocity profile of the original cluster are recovered by our algorithm.

A major goal for this study was to investigate what types of bias can occur when the GC's stars are sampled (a) randomly, (b) from the outer regions of the cluster, and (c) from the inner regions of the cluster. In summary, are findings are:

\begin{itemize}
    \item Using all spatial and kinematic information and sampling stars randomly from throughout the cluster, our method gives reliable credible intervals for the parameter values, as well as reliable cumulative mass profiles (CMPs),  and mean-squared velocity profiles.
    
    \item Using a biased sample of stars (i.e., within/outside $r_h$) gives unreliable credible intervals, leads to biased parameter estimates, and provides poor inference of the CMP and mean-square velocity profile.
    
    \item There are two possible exceptions where even biased samples still tend to be reliable: (1) extended and low $\Phi_0$ clusters that are sampled in the outer regions, and (2) compact clusters that are sampled in the inner regions. In these cases, we believe the credible intervals for the parameters and CMPs are more reliable because the distribution of the sampled data is more similar the true distribution of stars in the cluster. 

\end{itemize}

These results are quite promising.  If the stellar data is sampled randomly in an unbiased fashion, then our algorithm's estimates are quite accurate.  The mass profiles correspond closely to the theoretical curves, and the parameter estimates are close to the true parameters.  We are also able to accurately estimate our error range, so that our 50\% and 95\% credible regions for the parameters have very close to the correct coverage probabilities.

If the stars are instead sampled in a biased fashion, then the results are more mixed.  Biased sampling of outer stars only for an extended\textbf{ and low $\Phi_0$} cluster still works well, since the essential information is preserved.  However, in other cases, biased samples lead to biased estimates with poor coverage probabilities.  This is not surprising, since our model assumes that the star sample is truly random (i.e., unbiased). 

As we have seen, the bias in parameter estimates and profiles can be quite pronounced and consistent among the simulations when the data sample is biased. We could propose a ``calibration'' to correct for these parameter and profile biases, and such a calibration would allow us to re-scale the parameters and profiles to better match the truth. However, this calibration would only be valid for the specific analysis of full 6-D phase-space information that we have presented here. Ultimately, we plan to expand our method in future work to deal with projected position data and missing velocity components (i.e., a more realistic data scenario). At that stage, the biases in the mass and velocity profile estimates could change substantially. Thus, we leave any calibration to future work, when its application will be most useful.

There are many avenues to pursue for future work. We are currently investigating how to modify the model to give more accurate estimates in the face of biased samples, and similarly when only projected values of the star positions and velocities are known.

Both biased samples and missing data are an astronomer's reality. For example, kinematic data of stars measured by HST typically sample only a portion of the cluster, whereas the Gaia satellite mostly provides kinematic data from stars in a GC's outer regions with the inner regions being incomplete. Without accounting for a biased sample, parameter inference is less reliable. 

Real kinematic data from HST and Gaia also have well-understood measurement uncertainties. We have not included measurement uncertainties in our simulation study, but a valuable next step would be to generate noisy measurements and then include a measurement model for each star that takes into account the sampling distribution of the measured kinematic components. This step could be accomplished through a hierarchical model. Additionally, one could use this framework as a way to \textit{combine} data from different telescopes that have different measurement properties (e.g. HST and Gaia), and thus obtain a less-biased sample of the stars in the cluster. As we have shown in this work, an unbiased sample of stars is key to reliable parameter inference and recovering a good estimate of the CMP. 

Ultimately, astronomers are not only interested in the intrinsic properties of GCs, but are also interested in comparison and selection of GC models. The latter will help our understanding of internal GC dynamics and the larger story of GC evolution as GCs traverse the Galactic potential. For example, the recently developed \texttt{SPES} model \citep{Claydon2019} allows some of the stars in a GC to be ``potential escapers''. The existence of energetically unbound stars within clusters is, again, an astronomers reality and could strongly affect how well a given distribution function is fit to observations. In fact, \citet{deBoer2019} found that the \texttt{SPES} models were a better representation of Galactic GCs than \texttt{limepy} models when fitting to GC density profiles. We are currently investigating some preliminary model comparison tests with simulated data from the \texttt{limepy} and \texttt{spes} models (Lou et al, in prep).

It is also important to compare the method presented here to traditional methods in the literature that use the projected distances of stars to estimate density and mass profiles, and that combine data sets from different telescopes to use stars at all radii \citep[e.g.,][]{deBoer2019}. However, at this stage of our research we have assumed an ``ideal'' scenario in which we have the full 6-dimensional phase-space information of stars --- a comparison of our results to other methods which use only projected distances of the stars will unfairly favour our method simply because we have more positional information. In a follow-up study, we plan to improve our Bayesian approach so that it can be applied to the measurements of projected distances, and at this stage a more fair comparison of methods could be made.

The ability to attribute a given dynamical model to an observed GC is a key step towards unravelling a GCs current properties as well as its evolutionary history. Understanding the underlying distribution function of stars within a cluster allows for more complex GC features, like its dark remnant population, binary population, degree of mass segregation, and its tidal history to be more thoroughly explored. Using a model that incorporates all these components --- while also improving the statistical framework to account for sampling bias in observations --- will allow us to better understand the dynamical state of globular clusters. Knowing a cluster's dynamical state also places constraints on the cluster's properties at birth and how it has evolved over time. Hence, being able to fit a dynamical model to an observed GC strengthens the cluster's ability to be used as a tool to study the Universe around it.

\section*{acknowledgements}

GME acknowledges the support of a Discovery Grant from the Natural Sciences and Engineering Research Council of Canada (NSERC, RGPIN-2020-04554), and a Connaught New Reseacher grant from the University of Toronto. JSR was supported by NSERC grant RGPIN-2019-04142. JW would like to thank Mark Gieles for helpful discussions regarding the \texttt{limepy} software package. GME would like to thank Joshua Speagle for helpful discussions regarding the differential optimization algorithm. The authors would also like to thank the referee for their very helpful report that helped improve this paper.

\software{The code for this research can be found at \url{https://github.com/gweneadie/GCs}. Our code makes use of the following software and software packages: \texttt{astropy} \citep{2013A&A...558A..33A}, \texttt{Cairo} \citep{cairo}, \texttt{coda} \citep{coda}, \texttt{dplyr} \citep{dplyr}, \texttt{limepy} \citep{Gieles2015}, \texttt{MASS} \citep{MASS}, \texttt{NMOF} \citep{NMOF,NMOFbook}, \textbf{R} \citep{Rbase}, \texttt{reticulate} \citep{reticulate}, \texttt{tibble} \citep{tibble}, and \texttt{tidyverse} \citep{tidyverse}.}

\bibliography{EadieWebbRosenthal2021}{}
\bibliographystyle{aasjournal}

\end{document}